\def\eiso{E_{\rm iso}}

\def\eop{E^{\rm obs}_{\rm peak}}
\def\esp{E^{\rm src}_{\rm peak}}
\def\ep{E_{\rm peak}}

\documentclass[preprint2]{aastex}

\usepackage{color}

\shorttitle{Short GRB\,111117A}
\shortauthors{Sakamoto et al.}

\begin{document}


\title{Identifying the Location in the Host Galaxy of \\the Short GRB\,111117A with the {\it Chandra} Sub-arcsecond Position}


\author{
T.~Sakamoto\altaffilmark{1,2,3,4}, 
E.~Troja\altaffilmark{1,3,5,6},
K.~Aoki\altaffilmark{7}, 
S.~Guiriec\altaffilmark{5},
M.~Im\altaffilmark{8},
G.~Leloudas\altaffilmark{9,18},
D.~Malesani\altaffilmark{9},  
A.~Melandri\altaffilmark{10}, 
A.~de Ugarte Postigo\altaffilmark{9,13},
Y.~Urata\altaffilmark{11},
D.~Xu\altaffilmark{12},
P.~D'Avanzo\altaffilmark{10}, 
J.~Gorosabel\altaffilmark{13},
Y.~Jeon\altaffilmark{8},
R.~S\'{a}nchez-Ram\'{i}rez\altaffilmark{13},
M.~I.~Andersen\altaffilmark{9,19},
J.~Bai\altaffilmark{21,22},
S.~D.~Barthelmy\altaffilmark{3},
M.~S.~Briggs\altaffilmark{25},
S.~Foley\altaffilmark{26},
A.~S.~Fruchter\altaffilmark{15}, 
J.~P.~U.~Fynbo\altaffilmark{9},
N.~Gehrels\altaffilmark{3}, 
K.~Huang\altaffilmark{14},
M.~Jang\altaffilmark{8}, 
N.~Kawai\altaffilmark{16}, 
H. Korhonen\altaffilmark{19,24},
J.~Mao\altaffilmark{21,22,23},
J.~P.~Norris\altaffilmark{17},
R.~D.~Preece\altaffilmark{25},
J.~L.~Racusin\altaffilmark{3},
C.~C.~Th$\ddot{\rm o}$ne\altaffilmark{13},
K.~Vida\altaffilmark{20}
X.~Zhao\altaffilmark{21,22}
}

\altaffiltext{1}{Center for Research and Exploration in Space Science 
and Technology (CRESST), NASA Goddard Space Flight Center, Greenbelt, MD 
20771}
\altaffiltext{2}{Joint Center for Astrophysics, University of Maryland, 
	Baltimore County, 1000 Hilltop Circle, Baltimore, MD 21250}
\altaffiltext{3}{NASA Goddard Space Flight Center, Greenbelt, MD 20771}
\altaffiltext{4}{Department of Physics and Mathematics, College of Science and Engineering, Aoyama Gakuin University, 
5-10-1 Fuchinobe, Chuo-ku, Sagamihara-shi, Kanagawa 252-5258, Japan}
\altaffiltext{5}{NASA Postdoctoral Program Fellow, Goddard Space Flight Center, Greenbelt, MD 20771}
\altaffiltext{6}{Joint Center for Astrophysics, Department of Astronomy, University of Maryland
College Park, MD 20742-2421}
\altaffiltext{7}{Subaru Telescope, National Astronomical Observatory of Japan, 650 North A'ohoku Place, Hilo, HI 96720}
\altaffiltext{8}{Center for the Exploration of the Origin of the Universe (CEOU), Department of Physics and Astronomy, Seoul National University, Seoul, 151-747, Korea}
\altaffiltext{9}{Dark Cosmology Centre, Niels Bohr Institute, University of Copenhagen, Juliane Maries Vej 30, 2100 Copenhagen \O, Denmark} 
\altaffiltext{10}{INAF - Osservatorio Astronomico di Brera, via Bianchi 46, I-23807 Merate (LC), Italy}
\altaffiltext{11}{Institute of Astornomy, National Central University, Chung-Li 32054, Taiwan}
\altaffiltext{12}{Department of Particle Physics and Astronomy, The Weizmann Institute of Science, Rehovot 76100, Israel}
\altaffiltext{13}{Instituto de Astrof\'{i}sica de Andaluc\'{i}a (CSIC), Glorieta de la Astronom\'{i}a s/n, 18008 Granada, Spain}
\altaffiltext{14}{Academia Sinica Institute of Astronomy and Astrophysics, Taipei 106, Taiwan}
\altaffiltext{15}{Space Telescope Science Institute, 3700 San Martin Drive, Baltimore, MD 21218}
\altaffiltext{16}{Department of Physics, Tokyo Institute of Technology, 2-12-1 Ookayama, Meguro-ku, Tokyo 152-8551, Japan}
\altaffiltext{17}{Physics Department, Boise State University, 1910 University Drive, Boise, ID 83725}
\altaffiltext{18}{The Oskar Klein Centre, Department of Physics, Stockholm University, 106 91 Stockholm, Sweden}
\altaffiltext{19}{Niels Bohr Institute, University of Copenhagen, Juliane Maries Vej 30, 2100 Copenhagen, Denmark}
\altaffiltext{20}{Konkoly Observatory of the Hungarian Academy of Sciences, Konkoly Thege \'{u}t 15-17, 1121, Budapest XII, Hungary}
\altaffiltext{21}{Yunnan Astronomical Observatory, Chinese Academy of Sciences, Kunming, Yunnan Province, 650011, China}
\altaffiltext{22}{Key Laboratory for the Structure and Evolution of Celestial Objects, Chinese Academy of Sciences, Kunming, 650011, China}
\altaffiltext{23}{Space Science Division, Korea Astronomy and Space Science Institute, 776, Daedeokdae-ro, Yuseong-gu, Daejeon, 305-348, 
Republic of Korea}
\altaffiltext{24}{Centre for Star and Planet Formation, Natural History Museum of Denmark, University of Copenhagen, \O ster Voldgade 5-7, 
DK-1350 Denmark}
\altaffiltext{25}{Center for Space Plasma and Aeronomic Research, University of Alabama in Huntsville, 320 Sparkman Drive,
Huntsville, AL 35805, USA}
\altaffiltext{26}{Max-Planck-Institut f\"{u}r extraterrestrische Physik, Giessenbachstrasse 1, 85748 Garching, Germany}



\begin{abstract}

We present our successful {\it Chandra} program designed to identify, with sub-arcsecond 
accuracy, the X-ray afterglow of the short GRB\,111117A, which was discovered by {\it Swift} 
and {\it Fermi}.  Thanks to our rapid target of opportunity request, {\it Chandra} clearly detected 
the X-ray afterglow, though no optical afterglow was found in deep optical observations.  The 
host galaxy was clearly detected in the optical and near-infrared 
band, with the best photometric redshift of $z=1.31_{-0.23}^{+0.46}$ 
(90\% confidence), making it one of the highest known short GRB redshifts.    
Furthermore, we see an offset of $1.0 \pm 0.2$ arcseconds, 
which corresponds to $8.4 \pm 1.7$ kpc, 
between the host and the afterglow position.  
We discuss the importance of using {\it Chandra} for obtaining sub-arcsecond X-ray localizations  
of short GRB afterglows to study GRB environments.

\end{abstract}


\keywords{gamma rays: bursts}



\section{Introduction}

Gamma-ray bursts (GRBs) are traditionally divided in two 
classes based on their duration and spectral hardness: 
the long duration/soft spectrum GRBs, and the short duration/hard spectrum GRBs \citep{kouveliotou1993}. 
The two classes of bursts further differ in their spectral lags, 
the measurement of the delay in the arrival time of the low-energy photons
with respect to the higher energy ones: long bursts tend to have large positive lags, 
while short bursts exhibit negligible or negative lags \citep{norris2006}. 
The long standing paradigm is that these two phenomenological classes of GRBs originate 
from different progenitor systems. 
A preponderance of evidence now links long GRBs with the death of massive stars \citep[][and references therein]{woosley2006}, 
yet the origin of short GRBs remains largely unknown. 
The common notion that short bursts originate from coalescing compact binaries, 
either neutron star-neutron star (NS-NS) or neutron star-black hole (NS-BH)
mergers \citep[e.g.,][]{eichler1989,paczynski1991,narayan1992,rosswog2005,rezzolla2011}, 
makes them the most promising tool to aid in the direct detection of gravitational waves (GWs) 
by forthcoming facilities such as Advanced-LIGO, Advanced-VIRGO or LCGT (KAGRA) \citep[e.g.,][]{nissanke2010}.
It is therefore of primary importance to convincingly corroborate the merger 
scenario with a robust observational basis. 

Significant progress in understanding the origin of short GRBs has been achieved 
only recently. This advance was enabled by the detection of their afterglows in 
2005 thanks to the rapid position notice and response by {\it HETE-2} 
\citep{ricker2003} and {\it Swift} \citep{gehrels2004}.  
The very first localizations of short GRBs immediately provided us with fundamental clues about their nature. 
They demonstrated that short GRBs are cosmological events with an isotropic equivalent energy scale of 
10$^{49}$-10$^{52}$\,erg, 
that they occur in different environments than long GRBs, and are not associated with bright Type Ic supernovae 
\citep{bloom2006,prochaska2006,covino2006}. 

Since 2005 the sample of well-localized short GRBs has significantly grown, 
allowing for a deeper insight into the nature of their progenitors. 
The observed redshift distribution, ranging $0.11 \lesssim z \lesssim 1$,  
hints at a progenitor system with a broad range of lifetimes \citep{berger2007}.
Another critical test of the progenitor models is the observed offset distribution of short bursts 
\citep{troja2008,fong2010,church2011}. 
The median physical projected offset between the host center and the short GRB position 
is $\sim$5 kpc \citep{fong2010}, which is about five times larger than that of long GRBs \citep{bloom2002},
and shows a broader dispersion.
This is in agreement with the merger scenario, as several models
NS-NS/NS-BH systems are expected to receive significant kick velocities at birth \citep{bloom1999,fryer1999,belczynski2006}, 
or to dynamically form in globular clusters in the outskirts of their galaxies \citep{grindlay2006}.  

Despite the major progress of the last few years,  
the study of short GRBs and their progenitors has still been suffering from 
their less secure afterglow positions and redshifts.  
Unlike long GRBs, none of the redshifts of short GRBs
\footnote{Although the redshift of GRB 090426 (the $t_{90}$ duration 
measured by BAT is 1.24) has been measured from the absorption 
spectroscopy \citep{levesque2010,thone2011}, there are growing observational evidences 
that the progenitor of this GRB is akin to that of long GRBs  \citep[e.g.,][]{xin2011,lu2010}.
} 
has been directly measured 
through afterglow spectroscopy, and only in the case of GRB\,060121, a photometric redshift was
derived from the afterglow spectral energy distribution \citep{deUgartePostigo2006,levan2006}.  
This is because the optical afterglows are significantly fainter than those of long 
GRBs \citep{nysewander2009,kann2011}.
The redshifts of short GRBs are instead measured from spectroscopic observations of the 
``associated" host galaxy.  
The likelihood of a spurious association is small when a sub-arcsec position is available. 
However, if an afterglow is only detected by the {\it Swift} X-Ray Telescope \citep[XRT;][]{burrows},
the probability of a chance alignment is higher 
due to the larger uncertainty in the 
localization (2-5$^{\prime\prime}$). Unfortunately, the latter scenario represents the majority of cases ($\sim$65\% of the {\it Swift} short bursts sample).

A further bias is introduced by the fact that sub-arcsecond positions are mainly derived from optical afterglow detections, which are subject both to absorption along the line of sight and density effects.  
In fact, in the standard fireball model, the optical brightness depends sensitively 
on the density of the circumburst environment \citep{kp00}. This effect disfavors the accurate localization of short GRBs occurring in 
the lower-density galaxy halo or even outside their own galaxy, in the intergalactic medium. 
Such populations of large-offest short GRBs has already been suggested by \citet{bloom2007} and \citet{troja2008}. 
However being localized mainly by XRT, their association with the putative host galaxy remains uncertain.
Increasing the sample of large-offset short bursts with sub-arcsecond localization  is crucial to discriminate whether 
their progenitors were ejected from their birth site, favoring models which predict NS binaries with large kick velocities and $\sim$Gyr lifetimes, 
or they were formed from dynamical interactions in globular clusters \citep{salvaterra2010}.

In this context, rapid {\it Chandra} observations of short GRB afterglows represent
the critical observational gateway to overcome the current observational limits.
Since 65\% of {\it Swift} short GRBs are detected in X-rays, and only 25\% of them are 
detected in the optical band,  X-ray observations have a higher probability of detecting the afterglows
of short GRBs. 
The superb angular resolution of {\it Chandra} 
allows for a sub-arcsecond localization, comparable to optical localizations, 
thus enabling the secure host identification and the precise measurement of the GRB projected offset.  
Furthermore, because the X-ray afterglow is less subject to absorption and density effects, 
{\it Chandra} localizations allow us to build a sample of well-localized short GRBs with limited bias,
complementing the information derived from the sample of optically localized short GRBs.
This is the key to distinguish between the different possible short GRB populations \citep{sakamoto2009}, which could arise 
from a different progenitor and/or environment.


In this paper, we report the first results of our {\it Chandra} program which led to the 
accurate localization of GRB~111117A detected by {\it Swift} and {\it Fermi}.  
GRB~111117A is the 2nd short burst\footnote{The first one was GRB~111020A \citep{fong2012}.}   
in which the {\it Chandra} position is crucial for the host identification.  
Our results were leveraged with an intense ground-based follow-up campaign.  
No optical/infrared counterpart was found, therefore our {\it Chandra} localization
uniquely provides the only accurate sub-arcsecond position. 
The paper is organized as follows:  we introduce GRB 111117A in \S 2.    
In section \S 3, we describe the analysis softwares and methods used in this paper.  
We report the prompt emission properties in \S 4, 
the X-ray afterglow properties in \S 5.1, the deep optical afterglow limits in \S 5.2, and the host galaxy properties in \S 6.  
We discuss and summarize our results in \S 7.  The quoted errors are at the 90\% confidence level for  
prompt emission and X-ray afterglow data, and at the 68\% confidence level for optical and near infrared data
unless stated otherwise.  The reported optical and near infrared magnitudes are in the Vega system unless stated 
otherwise.  Throughout the paper, we use the cosmological parameters, $\Omega_{m}$ = 0.27, $\Omega_{\Lambda}$ = 0.73 
and $H_{0}$ = 71 km s$^{-1}$ Mpc$^{-1}$.  

\section{GRB\,111117A}

On 2011 November 17 at 12:13:41.921 UT, the {\it Swift} Burst Alert Telescope \citep[BAT;][]{barthelmy}
triggered and localized the short GRB\,111117A \citep{mangano2011}.  
The {\it Fermi} Gamma-ray Burst Monitor \citep[GBM;][]{meegan2009} also triggered on the burst 
\citep{foley2011}.  
The BAT location 
derived from the ground analysis was (R.A., Dec.) (J2000) = (00$^{\rm h}$ 50$^{\rm m}$ 49.4$^{\rm s}$ , 
+23$^{\circ}$ 00$^{\prime}$ 36$^{\prime\prime}$) with a 90\% error radius of 1.8$^{\prime}$.  The {\it Swift} XRT 
started its observation 76.8 s after the trigger.  
A fading X-ray source was found at the location of (R.A., Dec.) (J2000) = 
(00$^{\rm h}$ 50$^{\rm m}$ 46.22$^{\rm s}$ , +23$^{\circ}$ 00$^{\prime}$ 39.2$^{\prime\prime}$) with 
a 90\% error radius of 2.1$^{\prime\prime}$ \citep{melandri2011a}.  The {\it Swift} UV-Optical Telescope \citep[UVOT;][]{roming}  
began the observations of the field 137 s after the trigger, and no optical afterglow was detected \citep{oates2011}.  

The earliest ground observations of the field were performed by the Gao-Mei-Gu telescope 
(GMG) at 1.96 hr after the BAT trigger, and no afterglow was detected within the XRT error circle 
with an exposure time of 600 s in the $R$ band \citep{zhao2011}.  The Nordic Optical Telescope (NOT) observed the field 
at 8.9 hr after the burst, and found an optical source inside the XRT error circle \citep{andersen2011}, which   
was later confirmed to have a possible extended morphology by the Magellan/Baade  
telescope \citep{fong2011}, the Gemini-South telescope \citep{cucchiara2011}, the GROND telescope \citep{schmidl2011}, 
and the Telescopio Nazionale Galileo \citep[TNG;][]{melandri2011b}.  
The Gran Telescopio CANARIAS (GTC), the Subaru telescope, the United Kingdom Infrared Telescope 
(UKIRT) and the Canada-France-Hawai Telescope (CFHT) also collected images of the field.

Based on no clear detection of an optical afterglow of the short GRB\,111117A, we triggered our  
{\it Chandra} Target of Opportunity (ToO) observation 6 hr after the trigger \citep{sakamoto2011b}, 
and the observation started 3 days later.  The X-ray afterglow was clearly detected in 20 ks, 
obtaining a sub-arcsecond position of the afterglow in X-rays \citep{sakamoto2011c}.  

\section{Data Analysis}
HEAsoft version 6.11 and the {\it Swift} CALDB (version 20090130) were used for the {\it Swift} BAT 
data analysis.  The XRT data products were obtained from the automated results available from the UK 
{\it Swift} Science Data Center \citep{evans2007,evans2009}.
CIAO 4.3 and CALDB 4.4.6 were used for the {\it Chandra} data analysis.  
The {\it Fermi} Gamma-ray Burst Monitor (GBM) data were prepared using the 
RMFIT software package,\footnote{http://fermi.gsfc.nasa.gov/ssc/data/analysis/user/} 
with data from three Sodium Iodide (NaI) scintillation detectors (detector ID 6, 7 and 9) and 
two Bismuth Germanate (BGO) scintillation detectors (detector ID 0 and 1).  

A standard data reduction of optical and near infrared images was performed using the IRAF\footnote{http://iraf.noao.edu/} software
package.  SExtractor\footnote{http://www.astromatic.net/software/sextractor} \citep{bertin1996}, 
SkyCat Gaia\footnote{http://astro.dur.ac.uk/\~{}pdraper/gaia/gaia.html} 
and IRAF were used to extract sources and perform the photometry.  
To accomplish consistent photometry for images collected by various telescopes, we selected 
10 common stars in the field and performed relative photometry.  
%
When some of the stars 
were saturated (especially for a large aperture telescope such as GTC), a subset of these 10 reference stars were  
used.  The USNO B-1 R2 magnitude or the SDSS magnitudes were used as the reference magnitude 
for the stars.  For the near infrared images of UKIRT and CFHT, we use the reference stars in the 2MASS catalog.
The Galactic extinction has been corrected using $E(B-V) = 0.03$ mag toward the direction to this burst \citep{schlegel1998}.  
The log of optical and near infrared observations presented in this paper are summarized in Table \ref{tbl:obs_log}.  

\section{Prompt Emission}
\begin{figure}
\centerline{
\includegraphics[scale=0.4]{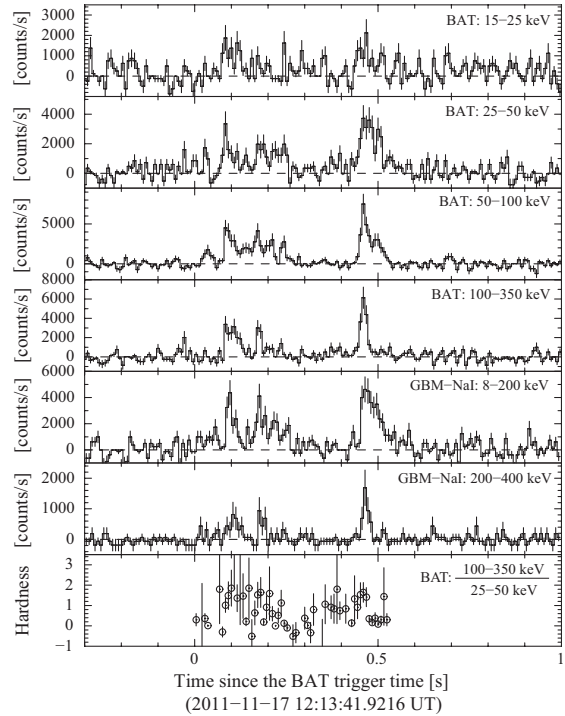}
}
\caption{The background subtracted 5 ms light curves of {\it Swift} BAT (15-25 keV, 25-50 keV, 50-100 keV and 100-350 keV) and {\it Fermi} 
GBM (8-200 keV and 200-400 keV).  The bottom panel shows the hardness ratio between the 100-350 keV and the 25-50 keV of the BAT 
data.  \label{bat_gbm_lc}}
\end{figure}

\begin{figure}
\centerline{
\includegraphics[scale=0.35]{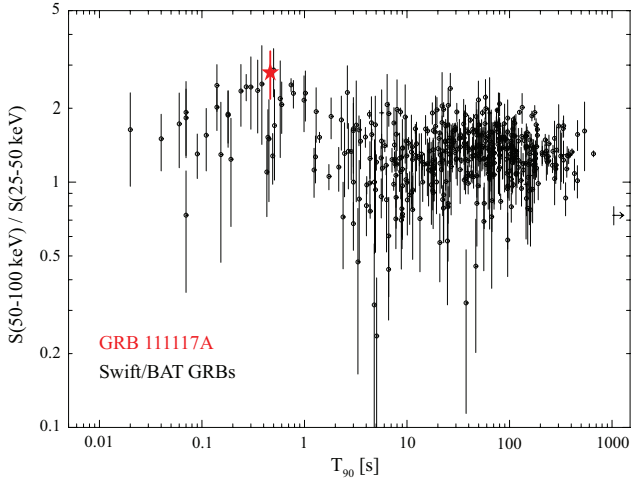}
}
\caption{Fluence ratios between the 50-100 keV and the 25-50 keV band versus $T_{90}$ are 
shown for GRB\,111117A (red) and the {\it Swift} BAT GRBs.  The values of the {\it Swift} BAT GRBs 
are extracted from \citet{sakamoto2011a}.
\label{bat_hard_t90}}
\end{figure}

The light curve of the prompt emission is composed of two episodes:  the first episode shows multiple overlapping 
pulses with a total duration of 0.3 s, and the second episode is composed of two pulses with 
a duration of 0.1 s (Figure \ref{bat_gbm_lc}).  
The duration is $T_{90}$ = $464 \pm 54$ ms (1 $\sigma$ error; 15-350 keV) measured using the BAT background-subtracted 
light curve using the mask modulation (e.g., mask-weighted light curve).  
This $T_{90}$ duration is significantly shorter than 2 s, which is the standard classification of short GRBs 
form BATSE \citep{kouveliotou1993}.  Furthermore, this duration is shorter than 0.7 s, 
which is claimed to be the dividing line between long and short 
GRBs for the {\it Swift} sample \citep{bromberg2012}. 
%
The hard-to-soft spectral evolution is 
seen in both the first and the second episode of GRB\,111117A (see the hardness ratio plot at the bottom panel 
of Figure \ref{bat_gbm_lc}).  There is no indication 
of extended emission \citep{norris2011} down to a flux level of $\sim$2 $\times$ $10^{-10}$ erg cm$^{-2}$ s$^{-1}$, 
assuming a power-law spectrum with a photon index of $\alpha = -2$ ($N(E) \propto E^{\alpha}$) in the 14-200 keV band by examining the BAT sky image from 
60 s (after the spacecraft slew settled) to 950 s after the BAT trigger time (hereafter $t_{0, \rm BAT}$).  
The spectral lag between the 100-350 keV and the 25-50 keV band is $0.6 \pm 2.4$ ms,  
which is consistent with zero, using the BAT raw light curves (non mask-weighted light curves) by subtracting 
a constant background measured around the burst.  
In the fluence ratio versus $T_{90}$ plane, GRB\,111117A is located in 
the same region where most of the BAT short GRBs are located (Figure \ref{bat_hard_t90}), 
further confirming its short GRB nature.   

The time-integrated spectral properties are investigated by performing a joint spectral analysis with 
BAT and GBM data.  The spectrum is extracted from $t_{0, \rm BAT}$ + 0.024 s to 
$t_{0, \rm BAT}$ + 0.520 s using {\tt batbinevt} for the BAT data and using the RMFIT software package 
for the GBM data in the same time interval.  
The BAT energy response file is generated by {\tt batdrmgen}.  
The GBM energy response files were retrieved from the HEASARC {\it Fermi} archive for trigger bn111117510.  
We use the {\tt xspec} spectral fitting package to do the joint fit.  The energy ranges of 15-150 keV, 8-900 keV and 0.2-45 MeV 
are used for the BAT, the GBM-NaI and the GBM-BGO instruments, respectively.  
The model includes a inter-calibration multiplicative factor to take into account the calibration uncertainty
among the different instruments.  
The best fit spectral parameters are summarized in Table \ref{tbl:prompt_spec}.  We find that a power-law multiplied by 
an exponential cutoff (CPL) \footnote{
$N(E) \propto E^{\alpha^{\rm CPL}} 
\exp\left(\frac{-E\,(2+\alpha^{\rm CPL})}{\ep}\right),$ 
where $\alpha^{\rm CPL}$ is the power-law photon index and $\ep$ is the
peak energy in the $\nu F_{\nu}$ spectrum.} provides the 
best representative model of the data.  The best fit parameters in this model are the power-law photon index 
$\alpha^{\rm CPL}$ = $-0.28_{-0.26}^{+0.31}$ 
and $E_{\rm peak}  = 420_{-110}^{+170}$ keV ($\chi^{2}$/d.o.f. = 627/661).  The 90\% confidence interval of 
the inter-calibration factor of the GBM detectors is between 0.50 and 0.78 which is an acceptable range taking into account 
the current spectral calibration uncertainty between the BAT and the GBM.  A simple power-law model yields a significantly 
worse fit to the data ($\chi^{2}$/d.o.f. = 729/662).  
Furthermore, the significant difference in the power-law photon index 
the BAT data ($-0.52_{-0.22}^{+0.24}$) and the GBM data 
($-1.44_{-0.08}^{+0.06}$) alone disfavors a simple power-law model as the representative model. 
There is no significant improvement in $\chi^{2}$ using 
a Band function \citep{band1993} fit ($\chi^{2}$/d.o.f. = 627/660) over 
a CPL fit.   The preferential fit to a CPL model and the systematically harder photon index compared to long GRBs 
are general characteristics of a time-integrated spectrum of short GRBs \citep[e.g.,][]{ghirlanda2009,ohno2008}.
The fluence in the 8-1000 keV band calculated using the best fit time-integrated spectral parameters based on a CPL fit 
above is $7.3_{-2.1}^{+2.6} \times 10^{-7}$ erg cm$^{-2}$.  Due to poor statistics in extracting a spectrum from a very short 
time window, the peak flux was calculated by scaling the BAT mask-weighted countrate into a flux by folding the BAT energy 
response and assuming the best fit time-integrated 
spectral parameters in a CPL model.  The peak energy flux at the 8-1000 keV band in the 50 ms window 
starting from $t_{0, \rm BAT}$ + 0.450 s is $(3.8 \pm 1.2) \times 10^{-6}$ erg cm$^{-2}$ s$^{-1}$.
The time-resolved spectroscopy is difficult to perform due to the limited statistics in the data.

We search for pre-burst emission by analyzing the BAT survey data (detector plane histogram; DPH).  
Approximately 4.5 hr before the burst trigger, GRB\,111117A was in the field of view of BAT 
(26.1$^{\circ}$ from the boresight direction) for $\sim$1 ks during 
the observation of the blazar PKS 0235+16 (observation ID 00030880085).  We use {\tt batsurvey} 
script to process the DPH data.  The extracted rates at the location of GRB\,111117A are corrected to 
the on-axis rate by applying an off-axis correction based on observations of the Crab.  We find no significant 
emission during this observation at the location of GRB\,111117A.  The 3 sigma upper limit assuming a 
power-law spectrum with a photon index of $\alpha=-2$ is $1.4 \times 10^{-9}$ erg cm$^{-2}$ s$^{-1}$ at the 14-200 keV band 
in 300 s exposure.

\section{Afterglow}

\subsection{X-rays}

\begin{figure}
\centerline{
\includegraphics[scale=0.35]{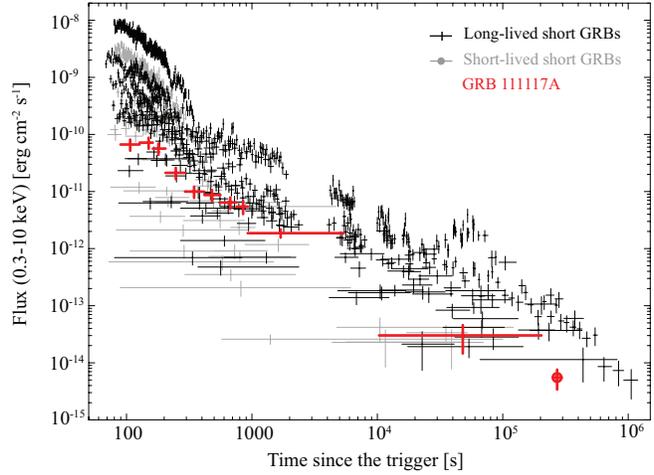}
}
\caption{Comparison of X-ray afterglow light curves of long-lived and short-lived short GRBs 
observed by {\it Swift} XRT and GRB\,111117A (red).  The {\it Chandra} data point of GRB\,111117A 
is shown in red filled circle.  The long-lived short GRBs include 
in this figure are GRB\,050724, GRB\,051221A, GRB\,051227, GRB\,060313, GRB\,061006, 
GRB\,061201, GRB\,061210, GRB\,070714B, GRB\,070724A, GRB\,070809, GRB\,071227, 
GRB\,080123, GRB\,080426, GRB\,090426, GRB\,090510, GRB\,090607, GRB\,090621B 
and GRB\,091109B.  The short-lived short GRBs include 
in this figure are GRB\,050509B, GRB\,050813, GRB\,051210, GRB\,060502B, GRB\,060801, 
GRB\,061217, GRB\,070429B, GRB\,080503, GRB\,080702A, GRB\,080905A, GRB\,080919, 
GRB\,081024A and GRB\,081226A.  
\label{xa}}
\end{figure}

The {\it Swift} XRT X-ray afterglow light curve can be fit to a simple power-law decay (Figure \ref{xa}).  
The spectrum collected in the photon counting (PC) mode is well described by an absorbed 
power-law model.  The best fit spectral parameters are a photon index of $-2.19_{-0.38}^{+0.36}$ 
and an excess $N_{H}$ of $1.8_{-1.0}^{+1.1} \times 10^{21}$ cm$^{-2}$ ($z=0$) assuming the galactic 
$N_{H}$ at the burst direction of $3.7 \times 10^{20}$ cm$^{-2}$ \citep{kalberla05}.  Both the measured photon index 
and $N_{H}$ for GRB\,111117A are consistent with those of other {\it Swift} short GRBs \citep{kopac2012,fong2012}.  

\begin{figure}
\centerline{
\includegraphics[scale=0.5]{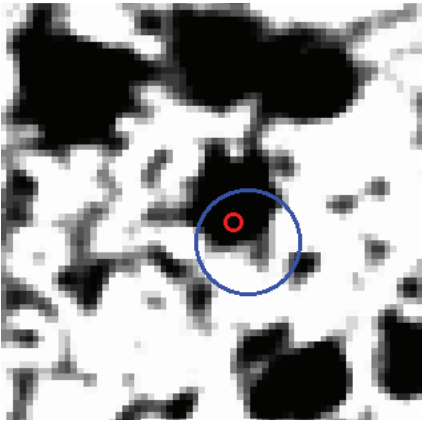}
}
\caption{GTC r image ($17^{\prime\prime} \times 17^{\prime\prime}$) with the XRT 90\% error circle in blue and the {\it Chandra} 
1 sigma error circle, which includes the statistical and the systematic error, in red.  \label{chandra_xrt_pos_gtc}}
\end{figure}

The {\it Chandra} observation started at 12:39:25 UT, and ended at 18:39:10  UT on 2011 November 20 
with a total exposure of 19.8 ks.  
The ACIS instrument had five CCD chips (S3, S4, S5, I2 and I3) enabled, with the S3 chip as the aiming point 
for the source.  The data were collected in the FAINT mode.  
The X-ray afterglow is clearly detected in the {\it Chandra} observation with 3.9 $\sigma$ significance by {\tt wavdetect} 
(source net counts of 8) within the XRT error circle.    To refine the astrometry of the {\it Chandra} data, 
we apply the same analysis method described in \citet{feng2008}.  We extract the {\it Chandra} image (0.35 - 8 keV) 
that overlaps with the GTC image ($4.4^{\prime} \times 8.7^{\prime}$).  
The astrometry of the GTC image is calibrated against the SDSS catalog, and its 
standard deviation is $\sim$0.3$^{\prime\prime}$.  We run {\tt wavdetect} with options of scales=``1.0 2.0 4.0 8.0 16.0" 
and sigthresh = $4 \times 10^{-6}$ to the extracted {\it Chandra} image.  There are four sources which 
have a good match between the images.  We then use the {\tt geomap} task in the IRAF IMMATCH package 
to find the best coordinate transformation between the {\it Chandra} and the GTC image by fitting those four sources.  
Finally, we apply {\tt geoxytrans} task (IRAF IMMATCH package) for the originally detected {\it Chandra} position 
using the coordinate transformation calculated by {\tt geomap} to find the astrometrically corrected {\it Chandra} 
afterglow position.  The refined afterglow position is shifted by $\delta$R.A. = $-0.221^{\prime\prime}$ and $\delta$Dec. 
= $-0.020^{\prime\prime}$ from the position originally derived by {\tt wavdetect}.  
The best {\it Chandra} X-ray afterglow position is 
(R.A., Dec.) (J2000) = (00$^{\rm h}$ 50$^{\rm m}$ 46.264$^{\rm s}$, +23$^{\circ}$ 00$^{\prime}$ 39.98$^{\prime\prime}$) 
with 1 $\sigma$ statistical uncertainty of 0.09$^{\prime\prime}$ in right ascension and 0.16$^{\prime\prime}$ in declination.  
When we include the systematic uncertainty of 0.3$^{\prime\prime}$, 1 $\sigma$ error radius of the 
{\it Chandra} position is 0.35$^{\prime\prime}$.  
The {\it Chandra} position is well within the XRT 90\% error circle (see Figure \ref{chandra_xrt_pos_gtc}).  

The combined {\it Swift} XRT and {\it Chandra} X-ray afterglow light curve is well fit by a simple 
power-law with index of $-1.25_{-0.12}^{+0.09}$.  As shown in Figure \ref{xa}, the X-ray afterglow 
of GRB\,111117A belongs to a dim population of the {\it Swift} short GRBs.  

\subsection{Optical}

\begin{figure}
\centerline{
\includegraphics[scale=0.3]{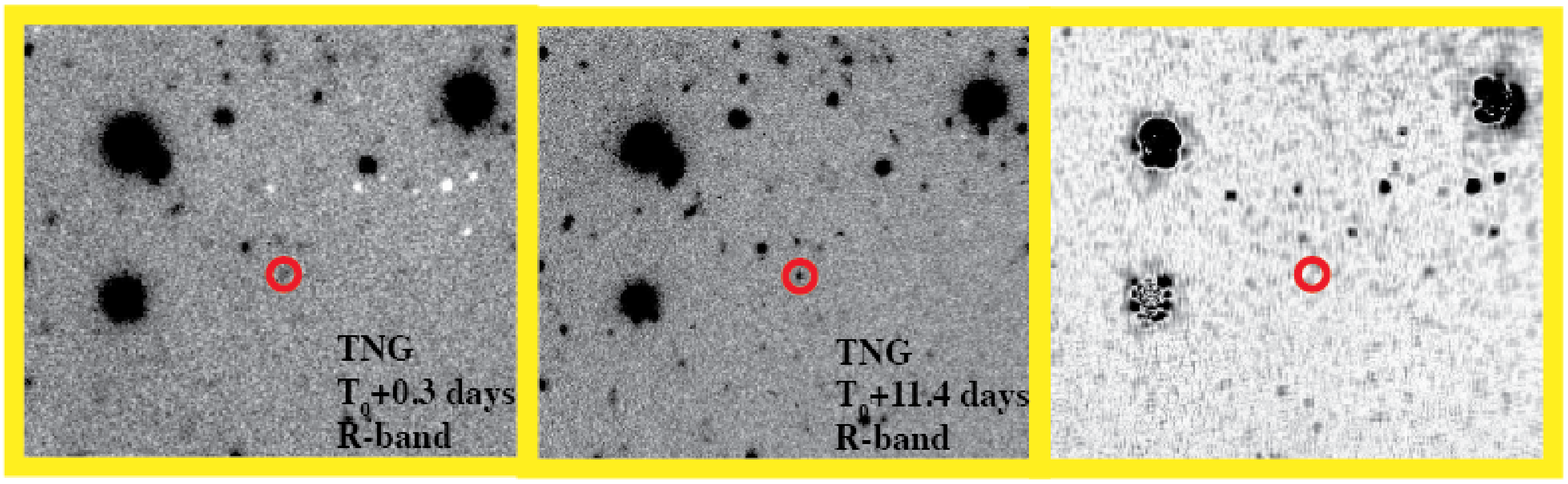}
}
\centerline{
\includegraphics[scale=0.2]{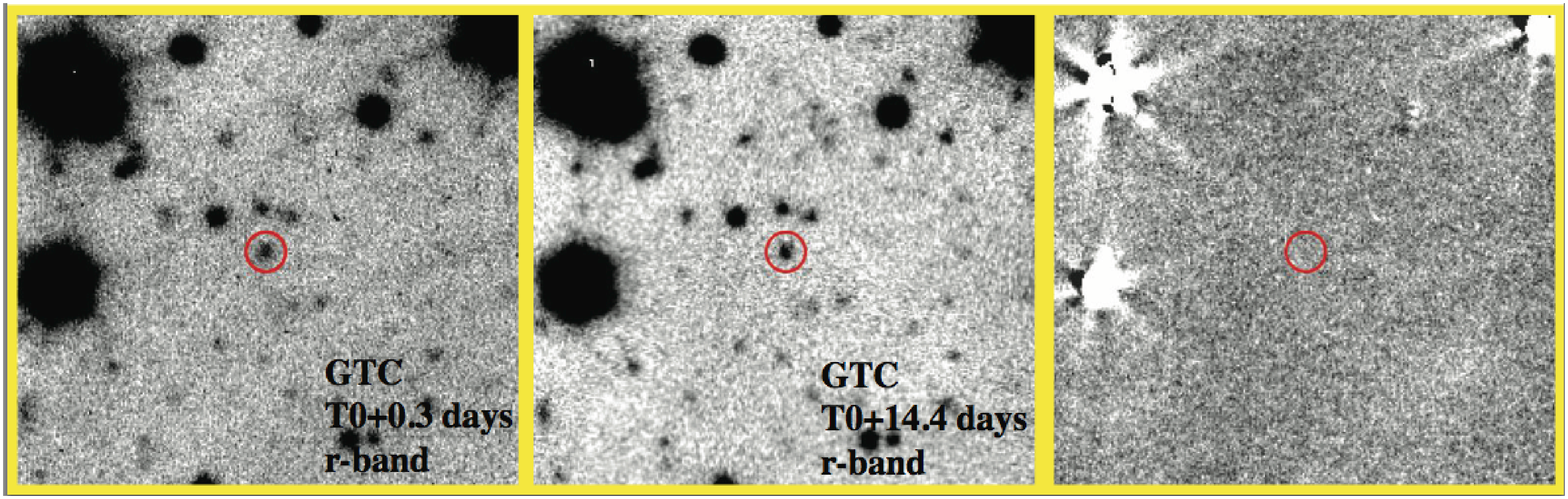}
}
\caption{Deep optical TNG (R; 1.4$^{\prime}$ $\times$ 1.2$^{\prime}$) and 
GTC (r; 1.1$^{\prime}$ $\times$ 1.0$^{\prime}$) images 
of two epochs.  The right 
panel shows the subtracted image of the first and second epoch.  No significant 
residuals are seen in both TNG and GTC subtracted images at the host location (red circle).  
\label{opt_sub}}
\end{figure}

\begin{figure}
\centerline{
\includegraphics[scale=0.35]{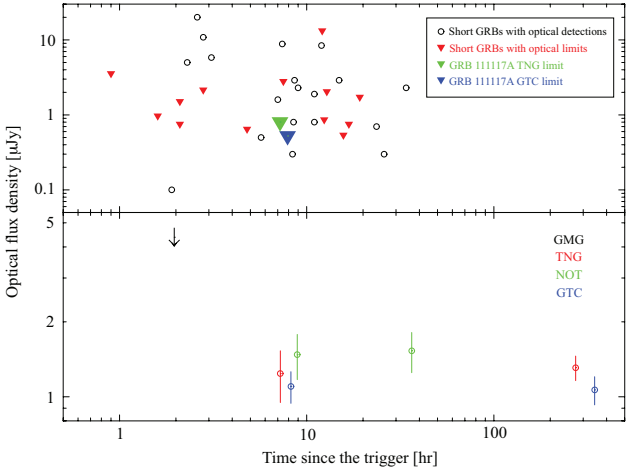}
}
\caption{Top: optical fluxes of the first optical detection (black circle) or an upper limit 
(filled triangle) of short GRBs \citep{berger2010} is shown as a function of the trigger time.  The TNG 
and GTC upper limits of the optical afterglow of GRB\,111117A are shown in green and blue 
filled triangle.  
Bottom: optical light curves of GRB\,111117A in $R$ and $r$ band are shown.  The plot includes 
$R$ band measurement from GMG, TNG and NOT, and also the $r$ band measurement from GTC.  
\label{fig:optical_ag_limit_optflux}}
\end{figure}

We investigate the possible optical afterglow emission by using the image subtraction technique 
between the early and the late time epoch observations by TNG and GTC.  We use the 
ISIS software package \citep{isis} to perform the image subtraction.  
The early and the late epoch observations of TNG and GTC were obtained at $t_{0, \rm BAT}$ + 7.23 
hr and  $t_{0, \rm BAT}$ + 7.89 hr, and  $t_{0, \rm BAT}$ + 11.4 days and $t_{0, \rm BAT}$ + 14.4 days, respectively.    
We find no significant emission at the {\it Chandra} X-ray afterglow location in the subtracted 
images in both the TNG and the GTC observations (Figure \ref{opt_sub}), 
with 3 sigma upper limits of $R$ $>$ 24.1 mag for TNG and $r$ $>$ 24.9 (AB) mag for 
GTC.  
%
The TNG limiting magnitudes of the first and second epochs are $R > 24.7$ mag and $R > 25.4$ 
mag, respectively.  For GTC, the limiting 
magnitude of the first and second epoch 
are $r > 25.8$ (AB) mag and $r > 26.1$ (AB) mag, respectively.  
Those limits are some of the deepest optical limits on short GRBs ever obtained (see upper panel of Figure 
\ref{fig:optical_ag_limit_optflux}).  

\section{Host Galaxy}

\begin{figure}
\centerline{
\includegraphics[scale=0.3]{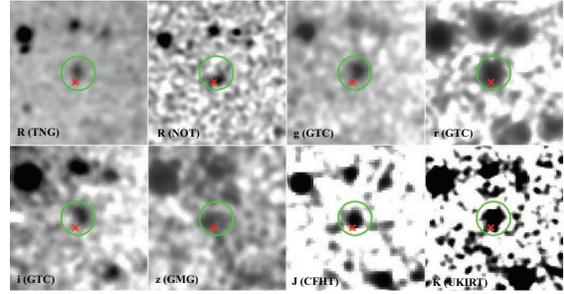}
}
\caption{Multi-color images at the field of GRB\,111117A.  From left to right, and top to bottom, 
the images are TNG $R$, NOT $R$, GTC $g$, GTC $r$, GTC $i$, GMG $z$, CFHT $J$ and UKIRT $K$.  The host 
galaxy is marked in a green circle.  The X-ray afterglow position determined by {\it Chandra} is marked as a red 
cross.  The image scale is 17$^{\prime\prime}$ $\times$ 17$^{\prime\prime}$.   
All the images are smoothed by the Gaussian function with 3 pixel radius.  \label{host_multi_image}}
\end{figure}

The host galaxy of GRB\,111117A has been detected in the near infrared and optical bands.  
%
%
There is only one near-infrared/optical source located near the {\it Chandra} X-ray 
afterglow position.  Although the weak nature of the source makes it difficult to investigate whether the 
source is extended or not, the optical flux of the source is constant between 7 hr and 
14 days after the burst at a level of $\sim$1.1 $\mu$Jy (bottom panel of Figure \ref{fig:optical_ag_limit_optflux}).  
Using the formula provided by \citet{bloom2002}, the probability of finding an unrelated 
galaxy of the $R$ magnitude of $\sim$23.3 with the distance of 1.0$^{\prime\prime}$ is 0.8\%.  
We also investigate the chance probabilities of the three nearby objects.  The probabilities of those 
objects are between 24\% and 42\% which are significantly larger than that of the host candidate.  Although 
the chance probability of the host candidate is non-negligible, the chance of a misidentification 
of the host galaxy is reasonably small.  
%
Therefore, we conclude that the source detected in the $K$, $J$, $z$, $i$, $r$, $g$, and $R$ bands is the host galaxy of GRB\,111117A 
(Figure \ref{host_multi_image}).  

\begin{figure}
\centerline{
\includegraphics[scale=0.4]{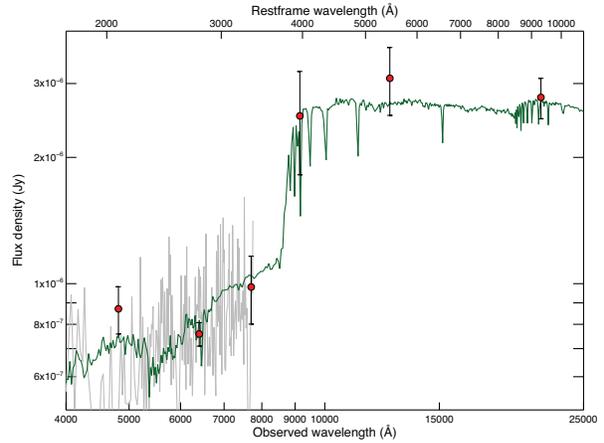}}
\vspace{0.5cm}
\centerline{
\includegraphics[scale=0.4]{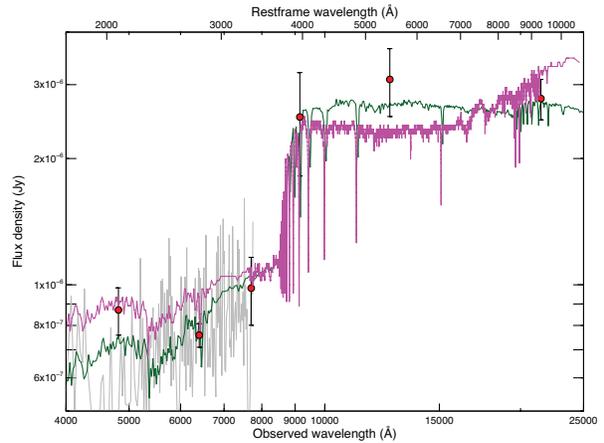}}
\caption{Top: The SED fit to the photometric data ($g$, $r$, $i$, $z$, $J$ and $K$) using the templates of the single stellar populations 
model \citep{maraston2005}. The GTC spectrum is shown in gray.  Bottom: The best fit SED template of the $\sim$0.1 Gyr 
post-starburst galaxy is overlaid in magenta.  \label{host_sed_bpz}}
\end{figure}

To estimate the redshift of the host galaxy, we perform a spectral energy density (SED) fit with the stellar population model of 
\citet{maraston2005}.  We use the single stellar populations (SSP) models with a Salpeter initial mass function 
\citep{salpeter1955}, solar metallicity which ranges from 0.005 $Z_{\Sun}$ to 3.5 $Z_{\Sun}$ (0.005, 0.02, 
0.5, 1.0, 2.0 and 3.5 $Z_{\Sun}$), and a red or blue horizontal branch morphology.  A total number of 269 SED templates 
ranging in stellar age from 10 Myr to 15 Gyr was applied.  The Bayesian Photometric Redshift 
software \citep[BPZ;][]{benitez2000} is used to fit the data in $g$, $r$, and $i$ bands (GTC), $z$ band (GMG), 
$J$ band (CFHT) and $K$ band (UKIRT) with those SED templates.  We find that the best fit SED template 
corresponds to a solar metallicity, a red horizontal branch morphology and the luminosity weighted mean 
stellar age of 0.1 Gyr with a 
redshift of 1.36$_{-1.18}^{+0.45}$.  Our best fit SED template of SSP model with a solar metallicity and a 
red horizontal branch matches well with other short GRB hosts studied by \citet{leibler2010}.  As seen 
in Figure \ref{redshift_prob_bpz} (top), there is a less significant low redshift solution ($z < 0.25$).  
We find that this low redshift solution is coming from the template with the young 
stellar age of $\sim$10 Myr.  As we will discuss in \S 7, it is unlikely that the host galaxy has the stellar 
age of $\sim$10 Myr.  Therefore, to constrain the 
redshift better, we focus on 41 SED templates with solar metallicity and a red horizontal branch morphology 
with stellar age from 20 Myr to 15 Gyr.  
The signal-to-noise is low, and there are no clear absorption or emission line 
features in the spectrum.  The continuum is consistent with the best fit SED template.  
The bottom panel of Figure \ref{redshift_prob_bpz} shows the posterior probability distribution of the estimated 
redshift for this SED fit.  No low redshift solution is evident in the probability distribution.  
We find the best estimated redshift to be 1.31 (90\% confidence interval $1.08 < z_{\rm ph} < 1.77$).  The likelihood 
that the redshift is correct is 80\% (reduced $\chi^{2}$ of the fit is 0.65 with 2 d.o.f.).  The best fit SED template is the case with the 
luminosity weighted mean stellar age of 0.1 Gyr and a mass of $\sim1 \times 10^{9}$ $M_{\odot}$.  
Figure \ref{host_sed_bpz} shows the best 
fit SED with the photometric data, and  the GTC spectrum with an exposure time of 4 $\times$ 1800 s.  
GTC spectroscopy was performed with the R1000B grism, which has a central wavelength of 5510 {\AA} and covers the spectral range between 3700 and 7000 {\AA} with a resolution of $\sim1000$ at 5500 {\AA}.
To investigate the likelihood of the host being a star-forming galaxy, we also examine the SED templates with an 
exponentially decaying star formation rate \citep{maraston2010}, with an $e$-folding time 
of 0.1, 1 and 10 Gyr (stellar age ranges from 10 Myr to 15 Gyr).  Although our $J$ band data point shows a relatively 
poor agreement with the best fit template with an $e$-folding time of 0.1 Gyr and the luminosity weighted mean stellar 
age of 0.3 Gyr, the fit is still acceptable 
(reduced $\chi^{2}$ = 0.91 with 2 d.o.f.).  
The best fit redshift in this case is 1.18$_{-0.21}^{+0.61}$.  The fit becomes worse if the $e$-folding time gets larger.  Therefore, 
our current data also support of an $\sim 0.1$ Gyr post-starburst galaxy (see bottom panel of Figure \ref{host_sed_bpz}).   
We also fitted the optical-NIR SED using MAGPHYS package
\citep{deCunha2008} to check the validity of the fitting result.
The MAGPHYS fit includes an extinction parameter as a part
of the fit, but performs the fitting at a given redshift. An exponentially
decaying star formation rate is assumed.  The redshifts were increased by 
a step of 0.05 from 0.8 to 1.5.  We confirm that the returned $\chi^2$ is the smallest in the range 
of the $z_{\rm ph}$ from the BPZ, and a moderate extinction of 
$A_{V} = 0.2-0.5$ mag is found.  The best fit solutions give
the exponential time scale of about 1.5 Gyr, with the luminosity 
weighted mean stellar age (in $r$-band) of a few hundred Myr,
and a stellar mass of a few times $10^{9}$ $M_{\odot}$.  These
output values are consistent with the solutions derived  
with the photometric redshift.  
In summary, 
based on various SED template fits, we can conclude the following about the host galaxy: 
the redshift 
of the host is $\sim$1.3 regardless of the SED model and the host is either a star-forming galaxy 
of the luminosity weighted mean stellar age of 0.1 Gyr and a mass of $\sim1 \times 10^{9}$ M$_{\odot}$ or 
a post-starburst galaxy.  
Further deep $J$ or $Y$ band data are crucial to pin down the host properties.  


A significant offset between the center of the host galaxy and the X-ray afterglow has been found 
for GRB\,111117A.  The center position of the host galaxy has been examined by running SExtractor on the second epoch 
of the GTC $r$ image, our highest quality optical image.  The best location of the host center is (R.A., Dec.) (J2000) 
= (00$^{\rm h}$ 50$^{\rm m}$ 46.258$^{\rm s}$, +23$^{\circ}$ 00$^{\prime}$ 40.97$^{\prime\prime}$).  
The position moves by less than half a pixel ($<0.13^{\prime\prime}$) by changing the detection 
threshold of SExtractor from 1.5 to 3.0 sigma.    Therefore, the projected offset between the center of the host galaxy and the 
X-ray afterglow is 1.0$^{\prime\prime}$ ($\delta$R.A. = $0.083^{\prime\prime}$ and $\delta$Dec. = $-0.990^{\prime\prime}$; 
see Figure \ref{host_multi_image}).  Taking into account the statistical error in the X-ray afterglow position 
of $0.18^{\prime\prime}$ and the statistical error of the host center location of $0.13^{\prime\prime}$, 
we estimate the offset with its error 
to be $1.0 \pm 0.2$ arcseconds, which corresponding to a distance of $8.4 \pm 1.7$ kpc at a redshift 
of  $z = 1.31$.  


\section{Discussion}

\begin{figure}
\centerline{
\includegraphics[scale=0.3]{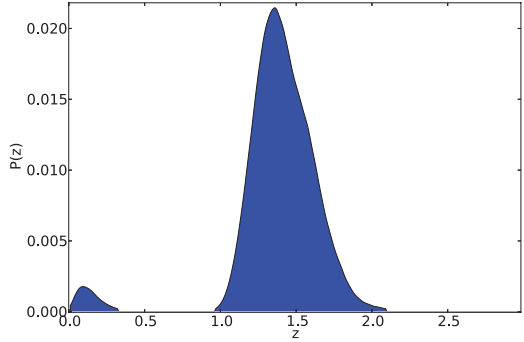}
}
\vspace{0.5cm}
\centerline{
\includegraphics[scale=0.3]{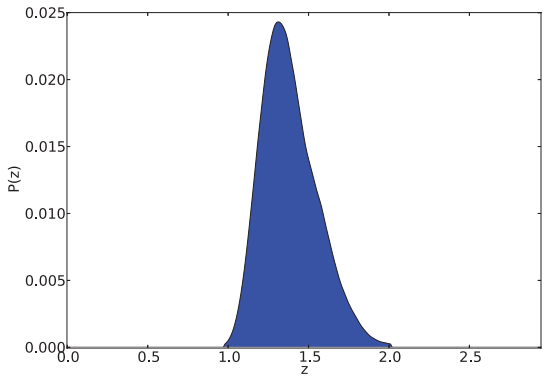}
}
\caption{Top: Posterior probability distribution of the photometric redshift by the SED fit of the host.  All 269 SED templates 
are used.  Bottom: Posterior probability distribution of the photometric redshift  of the host by 41 SED templates of a solar 
metallicity and a red horizontal branch morphology with stellar age from 20 Myr to 15 Gyr.  \label{redshift_prob_bpz}}
\end{figure}
\begin{figure}
\centerline{
\includegraphics[scale=0.35]{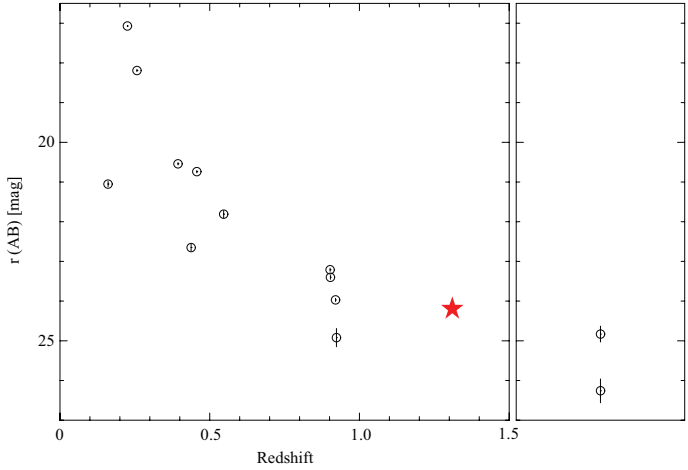}
}
\caption{Magnitude of the short GRB hosts as a function of redshift.  The right panel shows the magnitude 
of the hosts without a confirmed redshift.  GRB\,111117A (z=1.31) is shown as a red star.  
\label{host_z}}
\end{figure}

Our photometric redshift of $1.31_{-0.23}^{+0.46}$ (90\% confidence) 
for the host galaxy of GRB\,111117A is realistic for the following reasons.  
First, by plotting the observed $r$ 
AB magnitude ($r_{\rm AB}$) of the host galaxies of short GRBs as a function of redshift \citep{berger2009}, 
we find that the relatively faint magnitude of the host galaxy, $r_{\rm AB} = 24.20 \pm 0.07$, 
is located at the redshift range of $>$ 0.5 (Figure \ref{host_z}).  Second, 
we find that the less significant low redshift solution ($z < 0.25$) in the photometric redshift estimation (Figure 
\ref{redshift_prob_bpz}) is coming from the templates with the unrealistic young stellar ages of $\sim$10 Myr.    
At $z=0.25$, if the galaxy is star forming, there would be a chance of seeing emissions of 
[OII], H-$\beta$ and [OIII] in the optical spectrum, yet, we see none of those lines in the 
GTC spectrum.  
Furthermore, $\sim$10 Myr is in general too young for a whole galaxy, as opposed to a specific star 
forming region.  
Therefore 
the low redshift solution for the photometric redshift is unlikely to be the case of GRB\,111117A.  
Therefore, hereafter, we will discuss the rest-frame properties of GRB\,111117A using 
our best photometric redshift of 1.31.  

\begin{figure}
\centerline{
\includegraphics[scale=0.35]{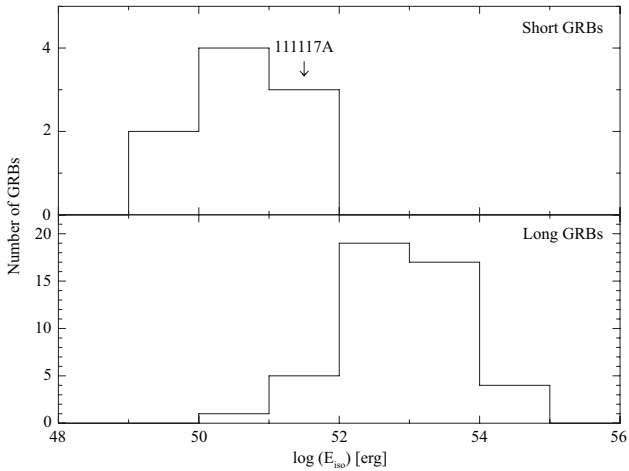}
}
\caption{Comparison of $\eiso$ between short (upper panel) and long (lower panel) GRBs.  
The short GRB $\eiso$ values are from \citet{berger2010} (short GRBs with detected afterglows and 
coincident host galaxies), and the long GRB $\eiso$ 
values are from \citet{nava2012} (only {\it Swift} long GRBs). 
GRB\,111117A ($3.4 \times 10^{51}$ erg) is located at the high end of $\eiso$ distribution of short GRBs.  
\label{eiso_hist}}
\end{figure}

Assuming the redshift of 1.31, the isotropic equivalent $\gamma$-ray energy ($E_{\gamma, \rm iso}$) 
which is integrated from 1 keV to 10 MeV in the rest frame is $3.4_{-1.5}^{+5.7}  \times 10^{51}$ erg.  
The peak energy at the rest-frame ($\esp$) is $945_{-310}^{+455}$ keV.  The 90\% errors in $E_{\gamma, \rm iso}$ 
and $\esp$ are taking into account not only a statistical error but also an uncertainty in the estimated redshift.  
As shown in Figure \ref{eiso_hist}, 
the $E_{\gamma, \rm iso}$ of GRB\,111117A is located at the high end of the $E_{\gamma, \rm iso}$ distribution 
of short GRBs and at the low end of the $E_{\gamma, \rm iso}$ distribution of long GRBs.  
Relatively low $E_{\gamma, \rm iso}$ and 
high $\esp$ compared to those of long GRBs make GRB\,111117A inconsistent with the $\esp$-$E_{\gamma, \rm iso}$ 
(Amati) relation \citep{amati2002}.  This characteristic is consistent with being a short GRB because most of the short GRBs 
are well known outliers of the Amati relation \citep{amati2006,nava2012}.  

The optical-to-X-ray spectral index \citep{jakobsson2004}, 
\begin{displaymath}
\beta_{\rm OX} (\equiv \log \{f_{\nu}(R)/f_{\nu}(\rm 3\, keV)\}/\log(\nu_{3\, keV}/\nu_{R})),
\end{displaymath} 
is estimated to be $\lesssim$ 0.78 
using the same definition on the X-ray flux density at 3 keV measured at 11 hrs after the burst, and the 
optical afterglow limit based on the GTC $r$ band.  
This upper limit of $\beta_{\rm OX}$ is within the allowed range of the standard afterglow model 
between 0.5 to 1.25.  Furthermore, 
according to \citet{margutti2012b}, the optical and the radio afterglow limit is consistent with the external shock 
model \citep{granot2002} for a small number density ($n$ $\lesssim$ 0.01-0.2 cm$^{-3}$).  However, there is a 
possibility that a significant amount of the optical afterglow flux was extinguished by the host galaxy.  
When we fit the 
X-ray afterglow spectrum to a power-law model with the intrinsic absorption at $z=1.31$, the 
intrinsic $N_{\rm H}$ is estimated to be $7.2_{-0.5}^{+0.7} \times 10^{21}$ cm$^{-2}$.  
Assuming a host extinction law similar to the Milky Way, $A_{V}$ is 4.1 mag \citep{predehl1995}.  Therefore, 
a significant amount of extinction 
in the optical flux is expected from the X-ray column density measurement.  On the other hand, it is still not clear 
whether it is possible to have such high extinction at the outskirts of the host where the X-ray afterglow 
is indicated.  Moreover, the amount of extinction which we derived from the SED fit of the host is $A_{V} = 0.2-0.5$ 
mag (see \S 6).  
%
%
At this stage, the origin of the large column density seen in the X-ray afterglow of 
GRB 111117A is still remains puzzling.  


The projected offset between the afterglow location and the host galaxy center is $8.4 \pm 1.7$ kpc 
using the estimated redshift of 1.31.  Although this offset is larger than the median projected offset of 
$\sim$5 kpc for previously studied short GRBs \citep{fong2010}, it is within the offset distribution of short GRBs.  
Using the projected offset of $r = 8.4$ kpc and the stellar age of $\tau = 0.1$ Gyr, the minimum kick velocity, 
$v = r/\tau$, is estimated to be $\approx$ 80 km s$^{-1}$.  The estimated kick velocity is similar to or possibly 
larger than the inferred kick velocity of GRB 060502B \citep{bloom2007}.  Using the typical age of 1-10 Gyr in the early-type 
short GRB hosts such as GRB 050509B, GRB 070809 and GRB 090515 \citep{bloom2006,berger2010}, 
the minimum kick velocity is estimated to be $\approx$ 1-8 km s$^{-1}$.  




In this paper, we have reported the prompt emission, the afterglow and the host galaxy properties 
of short GRB\,111117A.  The prompt emission observed by the {\it Swift} BAT and the {\it Fermi} GBM 
showed 1) a short duration, 2) no extended emission, 3) no measurable spectral lag and 4) a hard spectrum.  
All those properties can securely classify this burst as a short GRB.  
Although the optical afterglow 
has not been detected by our deep observations by TNG and GTC, our rapid {\it Chandra} ToO observation 
provides a sub-arcsec position of the afterglow in X-rays.  This {\it Chandra} position is crucial to identify 
the host galaxy and also to measure the significant offset of 1.0$^{\prime\prime}$ between the host center 
and the afterglow location.  
Our deep near infrared to optical photometry data of GMG, TNG, NOT, GTC, UKIRT and CFHT 
enable us to estimate the redshift of the host to 1.31.  
The observation of GRB\,111117A suggests that X-rays are more promising than optical to 
locate short GRBs with sub-arcsecond accuracy.  Combining the sub-arcsecond afterglow position 
in the X-ray and the deep optical images from the ground telescopes, we successfully investigate the host 
properties of GRB\,111117A even without an optical afterglow.  Rapid {\it Chandra} 
ToO observations of short GRBs are still key to increasing the {\it golden} sample of short GRBs with 
redshifts to pin down their nature.  

\acknowledgements
We would like to thank the anonymous referee for comments and suggestions that materially improved the paper.  
This work made use of data supplied by the UK {\it Swift} Science Data Centre at the University of Leicester.
This work is based on the observations using the United Kingdom Infrared Telescope, which is operated by the Joint 
Astronomy Centre on behalf of the Science and Technology Facilities Council of the U.K. with a partial support by {\it Swift} 
mission (e.g., {\it Swift} Cycle 7 GI grant NNX12AE75G) 
and the Gran Telescopio Canarias (GTC), installed in the Spanish Observatorio del Roque de los Muchachos of the Instituto 
de Astrof\'isica de Canarias, in the island of La Palma.  The Dark Cosmology 
Centre is funded by the Danish National Research Fundation. Partly based 
on observations made with the Nordic Optical Telescope, operated on the 
island of La Palma jointly by Denmark, Finland, Iceland, Norway, and 
Sweden, in the Spanish Observatorio del Roque de los Muchachos of the 
Instituto de Astrof\'isica de Canarias.  
This work was supported by {\it Chandra} Cycle 13 grant GO2-13084X.
The research activity of AdUP, CCT, RSR and JG is supported by Spanish research projects AYA2011-24780/ESP, 
AYA2009-14000-C03-01/ESP and AYA2010-21887-C04-01.  GL is supported by the Swedish Research Council through 
grant No. 623-2011-7117.  
KV is grateful to the Hungarian Science Research Program (OTKA) for support under the grant K-81421.  This work is 
supported by the "Lend\"ulet" Young Researchers' Program of the Hungarian Academy of Sciences.  
HK acknowledges the support from the European Commission under the Marie Curie IEF Programme in FP7.  
MI and YJ were supported by the he Creative Research 
Initiative program, No. 2010-0000712, of the National Research Foundation of Korea (NRFK) 
funded by the Korea government (MEST).   


\newpage

\newpage
\begin{deluxetable}{cccccccc}
\tabletypesize{\scriptsize}
\tablecaption{Log of optical and near-infrared observations of GRB\,111117A.  The magnitudes are corrected 
for Galactic extinction. \label{tbl:obs_log}}
\tablewidth{0pt}
\tablehead{
\colhead{Time since the trigger$^{\ast}$} &
\colhead{Telescope} &
\colhead{Instrument} &
\colhead{Filter} &
\colhead{Exposure} &
\colhead{Afterglow$^{\star}$} & 
\colhead{Host$^{\star}$} \\
\colhead{(day)} &
\colhead{} &
\colhead{} &
\colhead{} &
\colhead{(s)} &
\colhead{(mag)} &
\colhead{(mag)} &
}
\startdata
0.083 & GMG & YFOSC YSU & $R$ & 600 &  & $>22.1$ &\\  
0.104 & GMG & YFOSC YSU & $z$  & 900 &                & 22.9 $\pm$ 0.3 (AB)$^{\diamond}$\\ 
0.300 & TNG &  LRS  & $R$ & 1800 &  $>24.1$ & 23.49 $\pm$ 0.29\\
0.329 & GTC &  OSIRIS  & $g$  & 800   &  & 24.05 $\pm$ 0.14 (AB)\\
0.342 & GTC &  OSIRIS  & $r$  & 1200 &  $>24.9$ (AB) & 24.17 $\pm$ 0.10 (AB)\\
0.354 & GTC &  OSIRIS  & $i$  & 360   &   & 23.92 $\pm$ 0.20 (AB)\\
0.358 & NOT & ALFOSC & $z$ & 3600 & & $>22.5$ (AB) &\\
0.371 & NOT &  ALFOSC  & $R$ & 3000 &  & 23.20 $\pm$ 0.25\\  
1.5     & NOT &  ALFOSC  & $R$ & 2400 &  & 23.26 $\pm$ 0.22\\
11.4   & TNG &  LRS  & $R$ & 3600 &   & 23.43 $\pm$ 0.13\\
11.8   & Subaru & IRCS               & $K^{\prime}$ & 780 &  & $>19.95$\\ 
14.4   & GTC & OSIRIS  & $r$  & 2400 &   & 24.20 $\pm$ 0.07 (AB)\\
19 -- 72$^{\dagger}$  & UKIRT & WFCAM  & $K$  &  8640   &   &  20.91 $\pm$ 0.12 \\
42.7   & CFHT  & WIRCam          & $J$ &  4140 &  &  21.7 $\pm$ 0.2  \\
\enddata
\tablenotetext{\ast}{The mean time of the stacked observations with the corresponding exposure time.}
\tablenotetext{\dagger}{UKIRT data were collected at the multiple epochs between 19 and 72 days after the burst.}
\tablenotetext{\star}{Upper limit is in 3 sigma confidence level.}
\tablenotetext{\diamond}{Possible host detection.}
\end{deluxetable}

\newpage
\begin{deluxetable}{ccccccc}
\tablecaption{Time-integrated spectral parameters of GRB\,111117A.\label{tbl:prompt_spec}}
\tablewidth{0pt}
\tablehead{
\colhead{Instrument} &
\colhead{Model} &
\colhead{$\alpha$} &
\colhead{$\beta$} &
\colhead{$\eop$} &
\colhead{C(GBM)} &
\colhead{$\chi^{2}$/dof}
}
\startdata
BAT  & PL    & $-0.52_{-0.22}^{+0.24}$ & -- & -- & -- & 49.5/57 (0.87)\\\hline
GBM & PL    & $-1.44_{-0.08}^{+0.06}$ & -- & -- & -- & 636.8/604 (1.05)\\
GBM & CPL & $-0.39_{-0.37}^{+0.53}$ & -- & $440_{-125}^{+240}$ & -- & 578.9/603 (0.96)\\
GBM & Band & $-0.40_{-0.50}^{+0.36}$ & $<$$-2.6$ & $450_{-125}^{+240}$ & -- & 578.9/602 (0.96)\\\hline
BAT-GBM & PL & $-1.37_{-0.07}^{+0.05}$ & -- & -- & -- & 729.5/662 (1.10)\\
BAT-GBM & CPL & $-0.28_{-0.26}^{+0.31}$ & -- & $420_{-110}^{+170}$ & $0.62_{-0.13}^{+0.16}$ & 627.7/661 (0.95)\\
BAT-GBM & Band & $-0.28_{-0.26}^{+0.15}$ & $<$$-2.5$ & $420_{-80}^{+170}$ & $0.62_{-0.13}^{+0.16}$ & 627.8/660 (0.95)\\
\enddata
\end{deluxetable}

%

\newpage

\newpage

\newpage

\newpage

\newpage

\newpage

\newpage


\begin{thebibliography}{99}
\bibitem[Ackermann et al. (2010)]{ackermann2010} Ackermann, M., Asano, K., Atwood, W.~B., et al. 2010, ApJ, 716, 1178
\bibitem[Akerlof et al. (2011)]{akerlof2011} Akerlof, C.~W., Zheng, W., Pandey, S.~B., McKay, T.~A., 2011, ApJ, 726, 22
\bibitem[Alard \& Lupton (1998)]{isis} Alard, C., Lupton, R.~H. 1998, ApJ, 503, 325
\bibitem[Amati et al. (2002)]{amati2002} Amati, L., Frontera, F., Tavani, M., et al. 2002, A\&A, 390, 81
\bibitem[Amati (2006)]{amati2006} Amati, L. 2006, MNRAS, 372, 233
\bibitem[Andersen et al. (2011)]{andersen2011} Andersen, M.I., et al. 2011, GCN Circ. 12563, http://gcn.gsfc.nasa.gov/gcn3/12563.gcn3
\bibitem[Band et al. (1993)]{band1993} Band, D.~L., Matteson, J., Ford, L., et al. 1993, ApJ, 413, 281
\bibitem[Barthelmy et al. (2005a)]{barthelmy} Barthelmy, S.~D., Barbier, L.~M., Cummings, J.~R., et al.  2005a, 
\ssr, 120, 143
\bibitem[Barthelmy et al. (2005b)]{barthelmy2005b} Barthelmy, S.~D., Chincarini, G., Burrows, D.~N., et al. 2005b, Nature, 438, 994
\bibitem[Belczynski et al. (2006)]{belczynski2006} Belczynski, K., Perna, R., Bulik, T., et al. 2006, ApJ, 648, 1110
\bibitem[Ben\'itez (2000)]{benitez2000} Ben\'itez, N. 2000, ApJ, 536, 571
\bibitem[Berger et al. (2007)]{berger2007} Berger, E., Fox, D.~B., Price, P.~A., et al. 2007, ApJ, 664, 1000
\bibitem[Berger (2009)]{berger2009} Berger, E. 2009, ApJ, 690, 231
\bibitem[Berger (2010)]{berger2010} Berger, E. 2010, ApJ, 722, 1946
\bibitem[Berger (2011)]{berger2011} Berger, E. 2011, New Astronomy Reviews, 55, 1
\bibitem[Bernardini et al. (2012)]{bernardini2012} Bernardini, M.~G., Margutti, R., Zaninoni, E., Chincarini, G. 2012, MNRAS, 425, 1199
\bibitem[Bertin \& Arnouts (1996)]{bertin1996} Bertin, E., Arnouts, S. 1996, A\&AS, 117 393
\bibitem[Bloom et al. (1999)]{bloom1999} Bloom, J.~S., Sigurdsson, S., Pols, O.R. 1999, MNRAS, 305, 763
\bibitem[Bloom et al. (2002)]{bloom2002} Bloom, J.~S., Kulkarni, S.~R., Djorgovski, S.~G. 2002, AJ, 123, 1111
\bibitem[Bloom et al. (2006)]{bloom2006} Bloom, J.~S., Prochaska, J.~X., Pooley, D., et al. 2006, ApJ, 638, 354
\bibitem[Bloom et al. (2007)]{bloom2007} Bloom, J.~S., Perley, D.~A., Chen, H.-W., et al. 2007, ApJ, 654, 878
\bibitem[Bromberg et al. (2012)]{bromberg2012} Bromberg, O., Nakar, E., Piran, T., Sari, R. 2012, ApJ, 749, 110
\bibitem[Burrows et al. (2005)]{burrows} Burrows, S.~D., Hill, J.~E., Nousek, J.~A., et al.  2005, \ssr, 120, 165
\bibitem[Church et al.(2011)]{church2011} Church, R.~P., Levan, A.~J., Davies, M.~B., \& Tanvir, N.\ 2011, MNRAS, 413, 2004
\bibitem[Cobb et al.(2006)]{cobb2006} Cobb, B.~E., Bailyn, C.~D., van Dokkum, P.~G., \& Natarajan, P.\ 2006, ApJL, 651, L85
\bibitem[Covino et al.(2006)]{covino2006} Covino, S., Malesani, D., Israel, G.~L., et al.\ 2006, A\&A, 447, L5 
\bibitem[Cucchiara et al. (2011)]{cucchiara2011} Cucchiara, A., et al. 2011, GCN Circ. 12567, http://gcn.gsfc.nasa.gov/gcn3/12567.gcn3
\bibitem[D'Avanzo et al.(2009)]{davanzo2009} D'Avanzo, P., Malesani, D., Covino, S., et al.\ 2009, A\&A, 498, 711 
\bibitem[de Cunha et al. (2008)]{deCunha2008} de Cunha, E., Charlot, S., Elbaz, D., 2008, MNRAS, 388, 1595
\bibitem[de Ugarte Postigo et al. (2006)]{deUgartePostigo2006} de Ugarte Postigo, A., Castro-Tirado, A.~J., Guziy, S., et al. 2006, ApJ, 648, L83
\bibitem[Eichler et al. (1989)]{eichler1989} Eichler, D., Livio, M., Piran, T., Schramm, D.~M., 1989, Nature, 340, 126
\bibitem[Evans et al. (2007)]{evans2007} Evans, P.~A., Beardmore, A.~P., Page, K.~L., et al. 2007, A\&A, 469, 379
\bibitem[Evans et al. (2009)]{evans2009} Evans, P.~A., Beardmore, A.~P., Page, K.~L., et al. 2009, MNRAS, 397, 1177
\bibitem[Feng \& Kaaret (2008)]{feng2008} Feng, H., Kaaret, P. 2008, ApJ, 675, 1067
\bibitem[Foley et al. (2011)]{foley2011} Foley, S., et al. 2011, GCN Circ. 12573, http://gcn.gsfc.nasa.gov/gcn3/12573.gcn3
\bibitem[Fong et al. (2010)]{fong2010} Fong, W., Berger, E., Fox, D.~B. 2010, ApJ, 708, 9
\bibitem[Fong et al. (2011)]{fong2011} Fong, W., et al. 2011, GCN Circ. 12566, http://gcn.gsfc.nasa.gov/gcn3/12566.gcn3
\bibitem[Fong et al. (2012)]{fong2012} Fong, W., et al. 2012, ApJ, 756, 189
\bibitem[Fox et al. (2005)]{fox2005} Fox, D.~B., Frail, D.~A., Price, P.~A., et al. 2005, \nat, 437, 845
\bibitem[Fruchter et al. (2006)]{fruchter2006} Fruchter, A.~S., Levan, A.~J., Strolger, L., et al. 2006, \nat, 441, 4632 
\bibitem[Fryer et al. (1999)]{fryer1999} Fryer, C.~L., Woosley, S.~E., Hartmann, D.~H. 1999, ApJ, 526, 152
\bibitem[Granot \& Sari (2002)]{granot2002} Granot, J., Sari, R. 2002, ApJ, 568, 820
\bibitem[Gehrels et al. (2004)]{gehrels2004} Gehrels, N., Chincarini, G., Giommi, P., et al. 2004, ApJ, 611, 1005
\bibitem[Gehrels et al. (2005)]{gehrels2005} Gehrels, N., Sarazin, C.~L. O'Brien, P.~T., et al. 2005, \nat, 437, 851
\bibitem[Ghirlanda et al. (2009)]{ghirlanda2009} Ghirlanda, G., Nava, L., Ghisellini, G., Celotti, A., Firmani, C. 2009, A\&A, 496, 585
\bibitem[Gorosabel et al. (2006)]{gorosabel2006} Gorosabel, J., Castro-Tirado, A.~J., Guziy, S., et al. 2006, A\&A, 450, 87
\bibitem[Grindlay et al.(2006)]{grindlay2006} Grindlay, J., Portegies Zwart, S., \& McMillan, S.\ 2006, Nature Physics, 2, 116
\bibitem[Hjorth et al. (2005)]{hjorth2005} Hjorth, J., Watson, D. Fynbo, P.~U., et al. 2005, Nature, 437, 859
\bibitem[Jakobsson et al. (2004)]{jakobsson2004} Jakobsson, P., Hjorth, J., Fynbo, P.~U., et al. 2004, ApJ, 617, L21
\bibitem[Kalberla et al.(2005)]{kalberla05} Kalberla, P.~M.~W., Burton, W.~B., Hartmann, D., et al.\ 2005, \aap, 440, 775 
\bibitem[Kann et al. (2011)]{kann2011} Kann, D.~A., Klose, S., Zhang, B., et al. 2011, ApJ, 734, 96
\bibitem[Kopa{\v c} et al. (2012)]{kopac2012} Kopa{\v c}, D., et al. 2012, MNRAS, 424, 2392
\bibitem[Kouveliotou et al. (1993)]{kouveliotou1993} Kouveliotou, C., Meegan, C.~A., Fishman, G.~J., et al. 1993, ApJ, 413, L101
\bibitem[Kumar \& Panaitescu(2000)]{kp00} Kumar, P., \& Panaitescu, A.\ 2000, \apjl, 541, L51 
\bibitem[Levan et al. (2006)]{levan2006} Levan A.~J., Tanvir, N.~R., Fruchter, A.~S., et al. 2006, ApJ, 648, L9
\bibitem[Leibler \& Berger (2010)]{leibler2010} Leibler, C.~N., Berger, E. 2010, ApJ, 725, 1202
\bibitem[Levesque et al.(2010)]{levesque2010} Levesque, E.~M., et al. 2010, MNRAS, 401, 963
\bibitem[L$\ddot{\rm u}$ et al. (2010)]{lu2010} L$\ddot{\rm u}$, H., et al. 2010, ApJ, 725, 1965
\bibitem[Malesani et al.(2007)]{malesani2007} Malesani, D., Covino, S., D'Avanzo, P., et al.\ 2007, \aap, 473, 77
\bibitem[Mangano et al. (2011)]{mangano2011} Mangano, V., et al. 2011, GCN Circ. 12559, http://gcn.gsfc.nasa.gov/gcn3/12559.gcn3
\bibitem[Maraston (2005)]{maraston2005} Maraston, C. 2005, MNRAS, 362 799
\bibitem[Maraston (2010)]{maraston2010} Maraston, C., Pforr, J., Renzini, A., et al. 2010, MNRAS, 407, 830
\bibitem[Margutti et al. (2012a)]{margutti2012} Margutti, R., Zaninoni, E., Bernardini, M.~G., et al. 2012a, MNRAS, 428, 729
\bibitem[Margutti et al. (2012b)]{margutti2012b} Margutti, R., Berger, E., Fong, W., et al. 2012b, ApJ, 756, 63
\bibitem[Meegan et al. (1996)]{meegan1996} Meegan, C.~A., Pendleton, G.~N., Briggs, M.~S., et al. 1996, ApJS, 106, 65
\bibitem[Meegan et al. (2009)]{meegan2009} Meegan, C.~A., Giselher, L., Bhat, P.~N., et al. 2009, ApJ, 702, 791
\bibitem[Melandri et al. (2011a)]{melandri2011a} Melandri, A., et al. 2011a, GCN Circ. 12565, http://gcn.gsfc.nasa.gov/gcn3/12565.gcn3
\bibitem[Melandri et al. (2011b)]{melandri2011b} Melandri, A., et al. 2011b, GCN Circ. 12570, http://gcn.gsfc.nasa.gov/gcn3/12570.gcn3
\bibitem[Narayan et al. (1992)]{narayan1992} Narayan, R., Paczynski, B., Piran, T., 1992, ApJ, 395, L83
\bibitem[Nava et al. (2012)]{nava2012} Nava, L., Salvaterra, R., Ghirlanda, G., et al. 2012, MNRAS, 421, 1256
\bibitem[Nissanke et al. (2010)]{nissanke2010} Nissanke, S., Holz, D.~E., Hughes, S.~A., et al. 2010, ApJ, 725, 496
\bibitem[Nysewander et al. (2009)]{nysewander2009} Nysewander, M., Fruchter, A.~S., Pe'er, A. 2009, ApJ, 701, 824
\bibitem[Norris \& Bonnell (2006)]{norris2006} Norris, J.~P., Bonnell, J.~T. 2006, ApJ, 643, 266
\bibitem[Norris et al. (2011)]{norris2011} Norris, J.~P., Gehrels, N., Scargle, J.~D. 2011, ApJ, 735, 23
\bibitem[Oates et al. (2011)]{oates2011} Oates, S.~R., et al. 2011, GCN Circ. 12569, 12565, http://gcn.gsfc.nasa.gov/gcn3/12569.gcn3
\bibitem[Ohno et al. (2008)]{ohno2008} Ohno, M., Fukazawa, Y., Takahashi, T., et al. 2008, PASJ, 60, 361
\bibitem[Paczynski (1991)]{paczynski1991} Paczynski, B. 1991, Acta Astron., 41, 257
\bibitem[Predehl \& Schmitt (1995)]{predehl1995} Predehl, P., Schmitt, J. H. M. M, A\&A, 293, 889
\bibitem[Prochaska et al.(2006)]{prochaska2006} Prochaska, J.~X., Bloom, J.~S., Chen, H.-W., et al.\ 2006, ApJ, 642, 989 
\bibitem[Rezzolla et al.(2011)]{rezzolla2011} Rezzolla, L., Giacomazzo, B., Baiotti, L., et al.\ 2011, ApJL, 732, L6
\bibitem[Ricker et al. (2003)]{ricker2003} Ricker, G.~R., Atteia, J.-L., Crew, G.~B., et al. 2003, in AIP Conf. Proc. 662, Gamma-Ray Burst and Afterglow Astronomy 2001, ed. G.R.Ricker \& R.K. Vanderspek (New York: AIP), 3
\bibitem[Roming et al. (2005)]{roming} Roming, P.~W.~A., Kennedy, T.~E., Mason, K.~O., et al. 2005, \ssr, 120, 95
\bibitem[Rosswog(2005)]{rosswog2005} Rosswog, S.\ 2005, ApJ, 634, 1202
\bibitem[Sakamoto \& Gehrels (2009)]{sakamoto2009} Sakamoto, T., Gehrels, N. 2009, in AIP Conf. Proc. 1133, Gamma-Ray Bursts, 6th Huntsville Symposium, ed. C. Meegan, N. Gehrels \& C. Kouveliotou (Melville, New York), 112
\bibitem[Sakamoto et al. (2011a)]{sakamoto2011a} Sakamoto, T., et al. 2011a, ApJS, 195, 2
\bibitem[Sakamoto et al. (2011b)]{sakamoto2011b} Sakamoto, T., et al. 2011b, GCN Circ. 12562, http://gcn.gsfc.nasa.gov/gcn3/12562.gcn3
\bibitem[Sakamoto et al. (2011c)]{sakamoto2011c} Sakamoto, T., et al. 2011c, GCN Circ. 12580, http://gcn.gsfc.nasa.gov/gcn3/12580.gcn3
\bibitem[Salpeter (1955)]{salpeter1955} Salpeter, E.~E., 1955, ApJ, 121, 161
\bibitem[Salvaterra et al.(2010)]{salvaterra2010} Salvaterra, R., Devecchi, B., Colpi, M., \& D'Avanzo, P.\ 2010, MNRAS, 406, 1248
\bibitem[Sari et al.(1998)]{sari1998} Sari, R., Piran, T., \& Narayan, R.\ 1998, ApJL, 497, L17
\bibitem[Schlegel et al. (1998)]{schlegel1998} Schlegel, D.~J., Finkbeiner, D.~P., Davis, M. 1998, ApJ, 500, 525
\bibitem[Soderberg et al. (2006)]{soderberg2006} Soderberg, A.~M., Berger, E., Kasliwal, M., et al. 2006, ApJ, 650, 261
\bibitem[Schmidl et al. (2011)]{schmidl2011} Schmidl, S., et al. 2011, GCN Circ. 12568, http://gcn.gsfc.nasa.gov/gcn3/12568.gcn3
\bibitem[Th$\ddot{\rm o}$ne et al. (2011)]{thone2011}Th$\ddot{\rm o}$ne, C.~C., et al. 2011, MNRAS, 414, 479
\bibitem[Troja et al.(2008)]{troja2008} Troja, E., King, A.~R., O'Brien, P.~T., Lyons, N., \& Cusumano, G.\ 2008, MNRAS, 385, L10 
\bibitem[Villasenor et al. (2005)]{villasenor2005} Villasenor, J.~S., Lamb, D.~Q., Ricker, G.~R., et al. 2005, Nature, 437, 855
\bibitem[Woosley \& Bloom(2006)]{woosley2006} Woosley, S.~E., \& Bloom, J.~S.\ 2006, ARA\&A, 44, 507 
\bibitem[Xin et al. (2011)]{xin2011} Xin, L., et al. 2011, MNRAS, 410, 27
\bibitem[Zhao et al. (2011)]{zhao2011} Zhao, X.-H. et al. 2011, GCN Circ. 12560, http://gcn.gsfc.nasa.gov/gcn3/12560.gcn3
\bibitem[Zhang et al. (2009)]{zhang2009} Zhang, B., Zhang, B.-B., Virgili, F.~J., et al. 2009, 703, 1696
\bibitem[Zheng et al. (2012)]{zheng2012} Zheng, W., Akerlof, C.W., Pandey, S.B. et al. 2012, ApJ, 756, 64
\end{thebibliography}
\end{document}